# Human Mobility Patterns Modelling using CDRs

Suhad Faisal Behadili[1], Cyrille Bertelle[1]and Loay E. George[2]

[1]Normandie University, LITIS, FR CNRS 3638, ISCN, ULH, Le Havre, France
[2]Baghdad University, Computer Science Department, Baghdad, Iraq

## ABSTRACT

*The research objectives are exploring characteristics of human mobility patterns, subsequently modelling them mathematically depending on inter-event time Δt and traveled distances Δrparameters using CDRs (Call Detailed Records). The observations are obtained from Armada festival in France. Understanding, modelling and simulating human mobility among urban regions is excitement approach, due to itsimportance in rescue situations for various events either indoor events like evacuation of buildings or outdoor ones like public assemblies,community evacuation in casesemerged during emergency situations, moreover serves urban planning and smart cities. The results of numerical simulation are consistent withpreviousinvestigations findings in that real systems patterns are almost follow an exponential distribution.Further experimentations could classify mobility patterns into major classes (work or off) days, also radius of gyration could be considered as influential parameter in modelling human mobility, additionally city rhythm could be extracted by modelling mobility speed of individuals.*

## 1. INTRODUCTION

The studying of human mobility from mobile phone usage could be developed to resolve several issues like daily individuals activities, explain these activities, reveal the relationship between human mobility and his mobile usage, as well as exploit the feasibility of using mobile data to full understanding, and drawing human patterns precisely. The purpose of simulation and analysis is to acquire and analyze results in well conceptual vision. Hence, obtain high indications for decision makers, improve the city life patterns, sustainability, and to have active smart cities.Eventual real effects conditions and actions of specified system should be modeled mathematically,in order to explore the mobility characteristics [1, 2, 3, 4, 5]. However, they could be accomplished by pivoting on two kinds of events; the discrete events and frequent (continuous) events. By other hand, to model the uncertainty of system input variables, hence probability distribution and its parameters should be determined,which is the most critical and hardtopic in modeling and simulating any system. Many researches used CDRs to investigate collective and individual human mobility (Gonzalez et al., 2008), the segmentation of urban spaces (Reades et al. 2009), understand of social events (Calabrese et al., 2010). (Ratti et al., 2010) treat large telecommunication database for Great Britain to explore human interactions, withemphasizing on the highly interactions correlation among administrative regions. (Phithakkitnukoon et al., 2010) suggested the activity-aware map, using the user mobile to uncover the dynamic of inhabitants, for urban planning, and transportation purposes [10].

## 2. CASE STUDY DATA

Human behavior is a reflection of the digital traces of mobile usages; these traces could be considered social sensor data, which serve as a proxy for individuals'activities and mobility. The observed case study data are Call detail Records of mobile phone, consisting of 51,958,652 CDRs, represent entry activitiesof 615,712 subscribers, the observation period starting from Friday 4$^{th}$- Tuesday15$^{th}$of July in 2008.Figure 1 exploits the regular patterns of individuals' activities densities along total observed hours. It's clear that all days have similar pattern except day 9 due to lack of its data actually. In CDRs the individuals' occurrences in discrete (irrelevant) pattern only, means that any mobile individual's activity is recorded at (start/end) time;however there is a lack of information.Since, mobility tracing is supposed to have indicationfor the individual's occurrence either during active or inactive case (mobility with or without any mobile phone activity). There are no data meanwhile the mobile phone is idle, i.e. inactive or doesn't make any communication activities neither calls nor SMS activities. In addition, the only obtainable spatial data are the towers (X,Y) coordinates data, for that reasonthe individual's transitions from position to another (from tower to tower) would be estimated, therefore the positions would be resolved approximately with regarding to the tower coverage area (signal strength). With regarding to the non-deterministic anddiscrete nature of these data, accordingly the collective behavior would be an effective approach to be analyzed and simulated. In view of, each individual could be disappeared for a while from the DB.That's making individual tracing is unworthy with insignificant indications on individuals' behavior in the city [3].

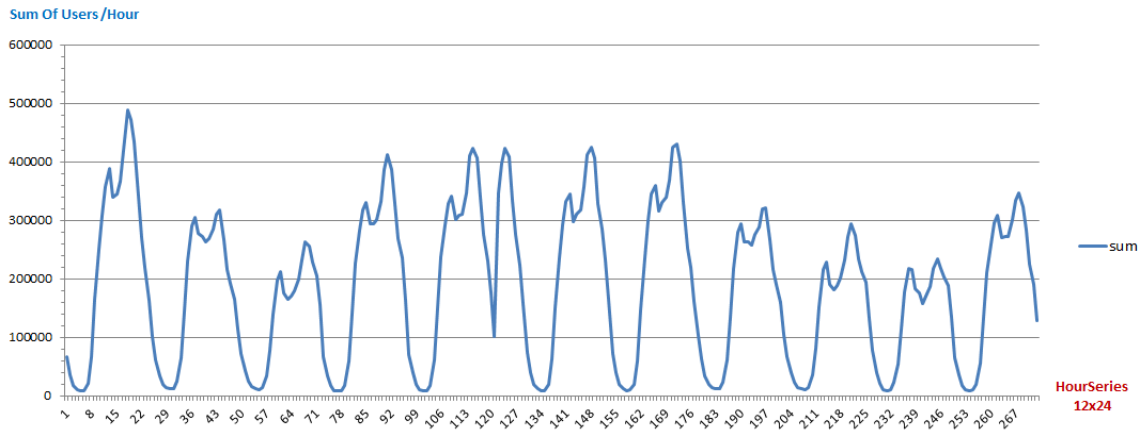

Figure 1: Activity patterns according to individuals' densities during days hours, over the observed area.

## 3. INDIVIDUALS' ACTIVITIES DESCRIBTION

The individuals are varied in their usage of mobile phone; hence their communications activities have heterogeneous nature. Theseactivities are ranged between (rarely-frequently) usages during specific period. Consequently, the individuals are grouped according to their total activities. The waiting time probability (inter-event time)$\Delta T$ of each two consecutive activities has been determined for each individual, and then individuals would be grouped according to their activities. Whereas, probability function has to be computed for capturing the system universal behavior (pattern), according to consecutive inter-event times $\Delta t_s$.Most of life systems aremodeled via exponential law [1, 6, 7, 5, 8]. So, the distribution of the average inert-event time $\Delta T$ is estimated by exponential distribution law as in equation (1), and its histogram is presented in figure 2. The demonstration in figure 3 exposesthose individuals of fewer activities have longer waiting times.

$$P(T) = (\Delta T)exp^{-\Delta T} \qquad (1)$$

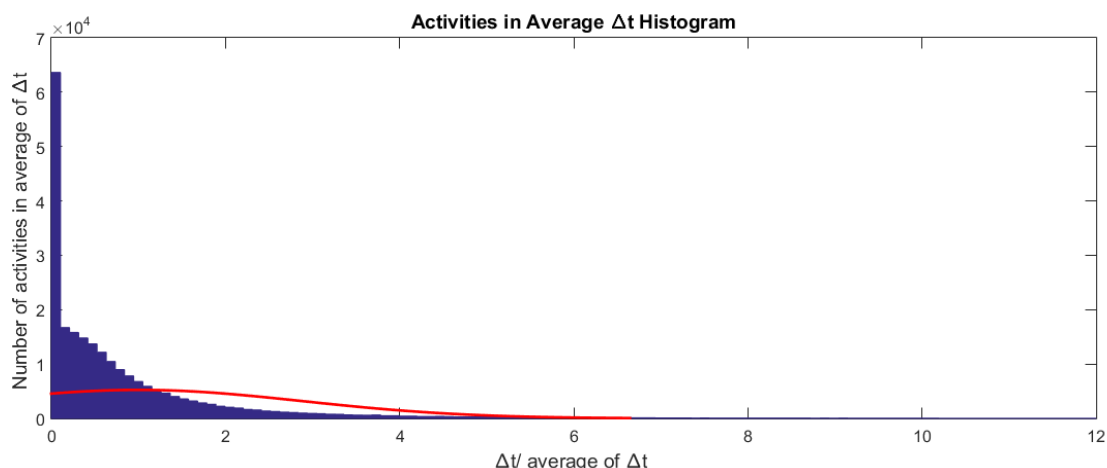

Figure 2: Activities in Average of Inter-Event TimeHistogram

## 4. OBSERVATIONS OF INTER-EVENT TIMES

The probability model computing represents starting step towards modeling and simulation process, which is modelling the data and simulating the model responses by using one of the well-known functions as the probability density function PDF [3,4, 9]. Here, in this study the mean of inter-event time is the essential parameter, whereas the means of samples of the distribution are drawn as exponential distribution. Thereafter compute the mean of all means to understand the sample means behavior from the exponential distribution. Consequently, get the universal system pattern (generalpopulation mobility law). The computations are done as follows:

1. Manipulate all 12 days data; each day is manipulated independently due to daily regular patterns, for all individuals in spatio-temporal form. Then eliminate the individuals of only one occurrence in the DB, since they have insignificant indication on mobility.
2. Sort the data by time to have the real sequence of positions transitions ofindividuals'trajectories. Thence classifying the data according to individuals' activities (sampling),by computing inter-event time ∆t (waiting time), where it is the time elapsed between each successive activities of each individual, which is here ranged between 15-1440 minutes.This sampling is done according to logical intuition, since 15 minutes is the minimum time that can give mobility indication, and the 1440 minutes (24 hours) is the highest elapsed time to travel inside the investigatedarea.
3. Compute the inter-event time ∆T and $\Delta T_a$ (average of inter- event time) for all individuals, then classify (min, max) samples according to activities score (activities densities), then compute $P(\Delta T)$.Where,$\Delta T/\Delta T_a$is the average inter-event time of all individuals (population).
4. Compute exponential distribution probability for each day, then for the whole days as in figure 3 to identify the universal population pattern law.As well as, figure 4 shows the approximation of all 12 days curves with the curve of their average values.

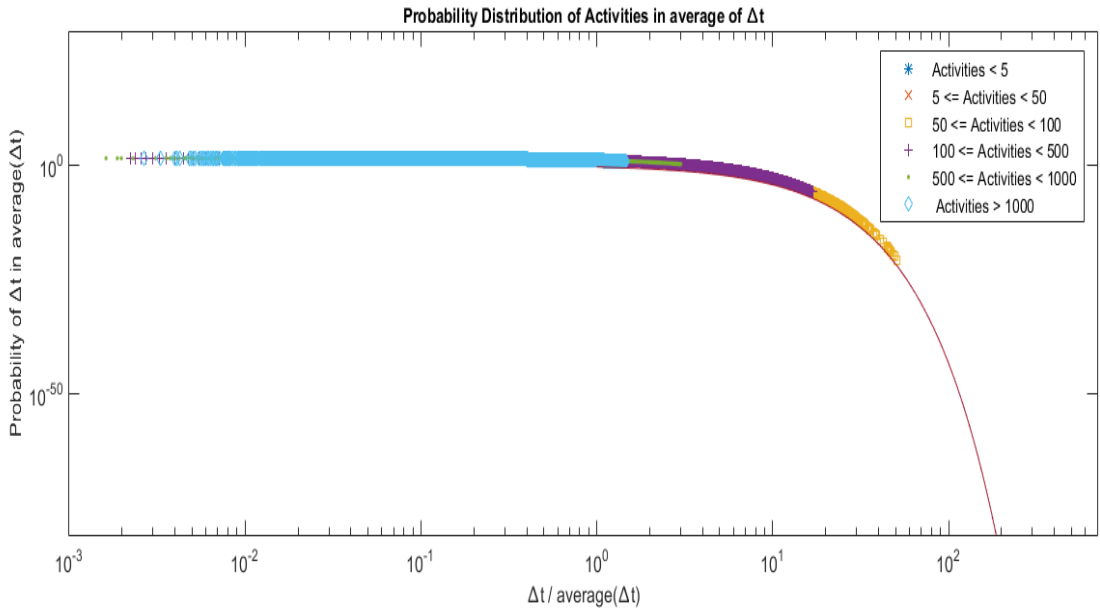

figure 3: Waiting time distribution P(Δt) of mobile activities, legend symbols distinguishes the individuals' groupsaccording to their activities ratio, for total population during totalperiod, activities inaverage Δt (spent time between each two successive activities).

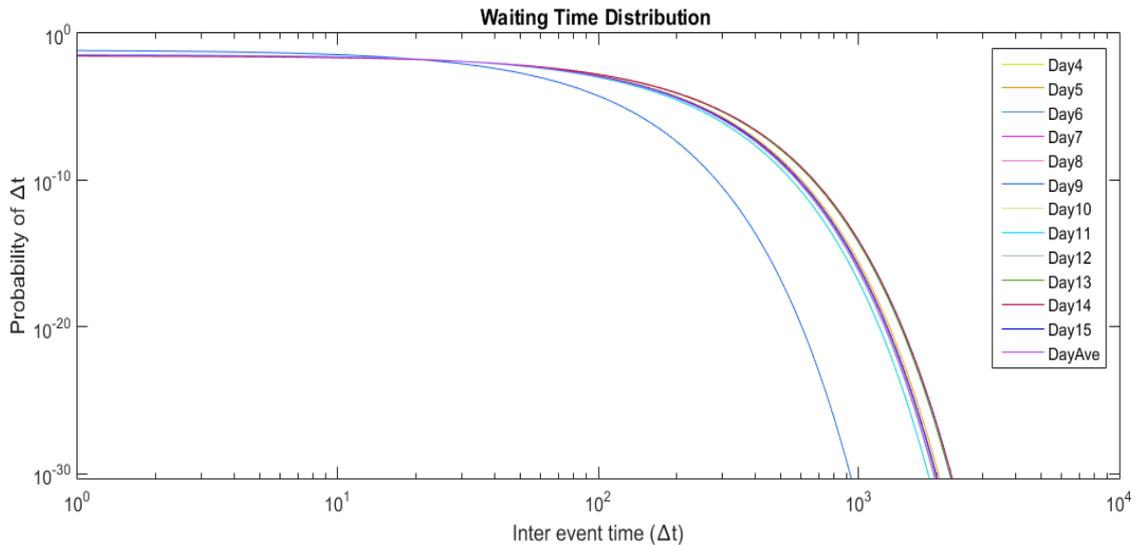

Figure 4: Waiting time distribution of the 12 days curves and the curve of their average (DayAve).

The exponential distribution (probability distribution) in equation (1) is capable of modeling the events that are happened randomly over time. Herein, it describes the inter-event time, and the average of inter-event time for individuals' activities [1, 2, 3, 9]. However, the cutoff distribution has beenfixedvia maximum observed inter-event time (significant parameters), which is the maximum time that individual can wait to make any mobile activity, where it is Δt=1431 min. The produced law as in equation (1), which is computed according to algorithm of complexity $O(n^3 + 2n^2 + n)$. The distribution exhibits that short inter-event times have higher probability than long ones, and the 12 days have similar patterns of activities. As well as, through following the same steps mentioned earlier to compute the displacement statistics of all individualsduring the whole observed period, the displacement probability distribution is computed. However, the Δr is the traveled distance between each two successive activities during time $\Delta T_0$ of the range 20-1440 minutes, whereP(Δr) is the displacement distribution.The investigated distances would be limited by the maximum distance that could be traveled by individuals in $\Delta T_0$ (time intervals). Hence, the cutoff distribution is determined by the maximum observed distance, which individual can travel is Δr=7.229515 Km along day hours, since the maximum time slice couldn't exceed 24 hours according to observed region. Displacements distribution is approximated by the power low as in equation (2), and the resulted distribution exhibits as in figure 5 that the displacements are clear in $10^4$m, whereas decreased after this value, where Δr is ranged between (0-1) Km.

$$P(\Delta r) = (\Delta r)exp^{-(\Delta r)} \qquad (2)$$

As demonstrated in figure 6 the distributionof traveled distance (displacements) for the 12 days curves in addition to the curve of their average, they have almost identical patterns. Hence, the obtained averages of the waiting time distribution are max ΔT = 1431 minute and min ΔT =0 minute, whereas the max Δr =7.229515e+04 meter and min Δr =0 meter.

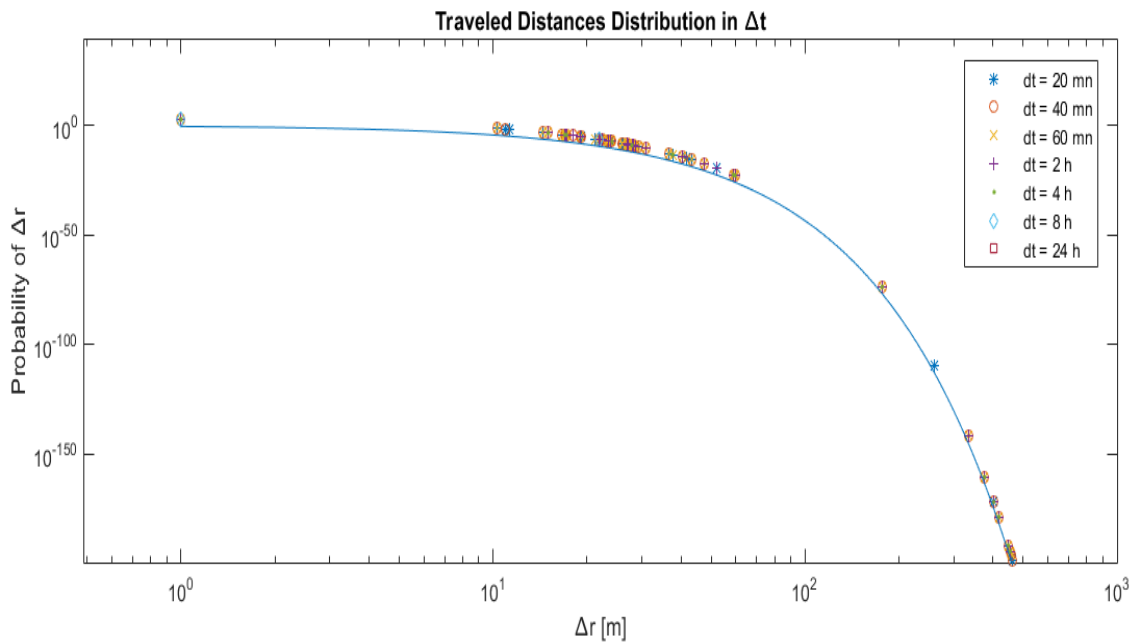

Figure 5: Distribution of distances probability (displacements) P(Δr) for waiting times (inter-event times) $\Delta t$ for one day, cutoff distribution is determined by the maximum distance traveled by individuals for specific $\Delta t_s$.

## 5. CONCLUSIONS

The catastrophic disasters, cities planning, diseases spreading, traffic forecasting, decision making for rescue people with disabilities in emergencies ...etc. they have impressive interest recently in various scientific fields [10]. All these activities require manipulating huge data and geographic data, in order to describe human behavior and mobility patterns, for giving the simulation model more reality, since almost simulation platforms are dealing with only a grid to be the background of the model, so the acquired model will be far away from reality.

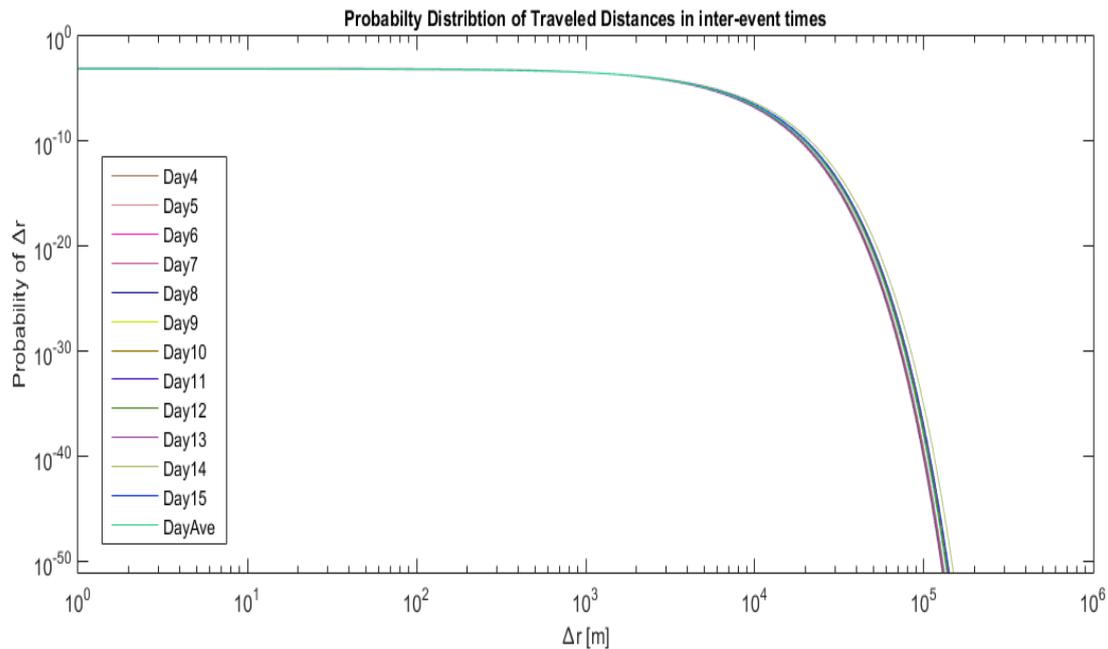

Figure 6: Displacements for the 12 days curves and the curve of their average (DayAve).

Based on this investigation the results showed that the inter-event time Δt between each two successive activities has bursty pattern, since there is long period without activities, this gives indication about the population's heterogeneity. As well as, human trajectories follow power-law distribution in step size Δr, and they are modelled using displacements Δr and waiting time Δt distributions.Additionally, the mobility traveling patterns show there are many short distances in contrast with few long distances.The population trajectories reconstruction could be performed using mobile phone data. Anywise, the results revealed that the patterns of all days are very similar and they have approximately identical curves either in spatial or temporal mode.Further research might explore the mobility patterns classificationsaccording to the (work or off) days for more understandings to the life patterns. Moreover, the radius of gyration could be considered as significant modelling parameter to give the model more reality with focusing on patterns regularity features.